\documentclass[a4paper,11pt]{article}
\usepackage{pos}

\title{Mixing and $C\!P$ violation in charm decays at LHCb}

\manuallySeparateAuthors
\author*{Federico Betti}
\author{ for the LHCb collaboration}

\affiliation{CERN (European Organization for Nuclear Research),\\
  Esplanade des Particules 1, 1211 Meyrin, Switzerland}

\emailAdd{federico.betti@cern.ch}

\abstract{The LHCb experiment has collected the world's largest sample of charmed hadrons. This sample is used to measure observables related to $D^0 -\overline{D}^0$ mixing, direct $C\!P$ violation and $C\!P$ violation in mixing and interference in the charm sector. In this document, the most recent results from LHCb on the search of direct $C\!P$ violation in $D^0$ and $D_{(s)}^+$ decays are summarised, as well as the first observation of mass difference between neutral charm-meson eigenstates and the most precise measurement of time-dependent $C\!P$ asymmetry in $D^0 \to K^+ K^-$ and $D^0 \to \pi^+ \pi^-$ decays.}

\FullConference{%
  *** The European Physical Society Conference on High Energy Physics (EPS-HEP2021), ***\\
  *** 26-30 July 2021 ***\\
  *** Online conference, jointly organized by Universität Hamburg and the research center DESY ***
}

\begin{document}
\maketitle

\section{Introduction}

For a generic hadron $M$ and its $C\!P$ conjugate $\overline{M}$, the decay amplitudes into the final state $f$ and its $C\!P$ conjugate $\overline{f}$ are defined as $A_f \equiv \langle f | \mathcal{H} | M \rangle$ and $\overline{A}_{\overline{f}} \equiv \langle \overline{f} | \mathcal{H} | \overline{M} \rangle$, where $\mathcal{H}$ is the effective decay Hamiltonian.
The $C\!P$ asymmetry of the decay $M \to f$ is defined as
\begin{equation}
    \mathcal{A}_{C\!P}(M \to f) \equiv \frac{\Gamma(M \to f) - \Gamma(\overline{M} \to \overline{f})}{\Gamma(M \to f) + \Gamma(\overline{M} \to \overline{f})} = \frac{1 - | \overline{A}_{\overline{f}}/A_f |^2}{1 + | \overline{A}_{\overline{f}}/A_f |^2}.
\label{ACP}
\end{equation}
When $ |\overline{A}_{\overline{f}} |^2 \neq |A_f |^2$, direct $C\!P$ violation takes place in the decay.
For neutral mesons such as $D^0$, the eigenstates $|D_H\rangle$ and $|D_L\rangle$ of the effective Hamiltonian, which have defined masses ($m_H$ and $m_L$) and decay widths ($\Gamma_H$ and $\Gamma_L$), can be written as a superposition of flavour eigenstates $|D_{L,H}\rangle = p|D^0\rangle \pm q|\overline{D}^0 \rangle$, where $p$ and $q$ are complex numbers satisfying $|p|^2 + |q|^2 = 1$.
The mixing is described by the parameters
\begin{align}
    x &\equiv \frac{m_H - m_L}{\Gamma}, \\
    y &\equiv \frac{\Gamma_H - \Gamma_L}{2\Gamma},
\label{mix_pars}
\end{align}
with $\Gamma = (\Gamma_H + \Gamma_L) / 2$.
The time-dependent $C\!P$ asymmetry between the probability of a meson produced in the $D^0$ and $\overline{D}^0$ flavour eigenstates at time 0 to decay into the final state $f$ after a time $t$
\begin{equation}
    \mathcal{A}_{C\!P}(D^0(t) \to f) \equiv \frac{\Gamma(D^0(t) \to f) - \Gamma(\overline{D}^0(t) \to f)}{\Gamma(D^0(t) \to f) + \Gamma(\overline{D}^0(t) \to f)}
\label{ACP_td}
\end{equation}
differs from 0 if $C\!P$ violation occurs either in the mixing (when $|q/p| \neq 1$) or in the interference between mixing and decay (when $\phi_f \equiv \operatorname{arg}[ (q \overline{A}_f) / (p A_f) ] \neq 0$).
In the $D^0$ meson system, the mixing parameters are predicted to be $\mathcal{O}(10^{-3})$~\cite{Falk:2001hx,Gronau:2012kq,Cheng:2010rv,Jiang:2017zwr,Wolfenstein:1985ft,Donoghue:1985hh}, and the time-dependent $C\!P$ asymmetry can be written as
\begin{equation}
    \mathcal{A}_{C\!P}(D^0(t) \to f) \approx \mathcal{A}_{C\!P}(D^0 \to f)+ \frac{4 |A_f|^2 |\overline{A}_f|^2}{(|A_f|^2+|\overline{A}_f|^2)^2} \Delta Y_f \frac{t}{\tau_{D^0}},
\label{ACP_td_approx}
\end{equation}
where $\tau_{D^0}$ is the $D^0$ lifetime and $\Delta Y_f$ is the observable describing the amount of time-dependent $C\!P$ violation~\cite{Grossman:2009mn,Kagan:2009gb}.
The coefficient in front of $\Delta Y_f$ differs from unity by $\mathcal{O}(10^{-6})$, resulting in $\Delta Y_f$ being equal to the slope of $\mathcal{A}_{C\!P}(D^0(t) \to f)$.

In the Standard Model (SM) the $C\!P$ asymmetries in charm decays are expected to be of the order of $10^{-4}$--10$^{-3}$.
Due to the presence of low-energy strong-interaction effects, theoretical predictions are difficult to compute reliably~\cite{Golden:1989qx,Buccella:1994nf,Bianco:2003vb,Artuso:2008vf,Brod:2011re,Cheng:2012wr,Cheng:2012xb,Li:2012cfa,Franco:2012ck,Pirtskhalava:2011va,Feldmann:2012js,Brod:2012ud,Hiller:2012xm,Grossman:2012ry,Bhattacharya:2012ah,Muller:2015rna,Khodjamirian:2017zdu,Buccella:2019kpn}.

The measurement of $\Delta \mathcal{A}_{C\!P} \equiv \mathcal{A}_{C\!P}(D^0 \to K^+ K^-) - \mathcal{A}_{C\!P}(D^0 \to \pi^+ \pi^-)$ with the full data sample collected by the LHCb detector led to the first observation of $C\!P$ violation in the decay of charm hadrons~\cite{Aaij:2019kcg}.
The result challenges perturbative estimates of $\Delta \mathcal{A}_{C\!P}$~\cite{Grossman:2006jg,Khodjamirian:2017zdu} and therefore prompted a renewed interest of the theory community in the field, sparking a discussion whether the measured value is consistent with the SM or if it is a sign of new physics~\cite{Chala:2019fdb,Grossman:2019xcj,Li:2019hho,Soni:2019xko,Cheng:2019ggx,Dery:2019ysp,Wang:2020gmn,Bause:2020obd,Dery:2021mll,Cheng:2021yrn,Kagan:2020vri}.
Additional measurements of $C\!P$ violation in the charm sector are thus crucial to clarify the picture and solve open theoretical puzzles.

In this document, the most recent results obtained by the LHCb collaboration in the field of $C\!P$ violation in charm are summarised.

\section{Measurements of direct $C\!P$ asymmetries in $D^0$ and $D_{(s)}^+$ decays}

The LHCb collaboration has recently published two measurements of direct $C\!P$ asymmetry in the decay of charm mesons.
The first one is the measurement of $C\!P$ asymmetry of the $D^0 \to K_s^0 K_s^0$ decay.
The only contributing amplitudes in the $D^0 \to K_s^0 K_s^0$ decay proceed via tree-level exchange and loop-suppressed diagrams of similar size which vanish in the flavour-SU(3) limit, and their interference could result in a $C\!P$ asymmetry up to $\mathcal{O}(10^{-2})$ in the SM~\cite{Nierste:2015zra}.
The result of the measurement, performed with data collected during the LHC Run~2, corresponding to an integrated luminosity of $6~\mathrm{fb}^{-1}$, is
\begin{equation*}
    \mathcal{A}_{C\!P}(D^0 \to K_s^0 K_s^0) = (-3.1 \pm 1.2 \pm 0.4 \pm 0.2)\%,
\end{equation*}
where the first uncertainty is statistical, the second systematic and the third is related to the knowledge of $\mathcal{A}_{C\!P}(D^0 \to K^+ K^-)$~\cite{Aaij:2021uvv}.
This is the most precise determination of this quantity to date, it is in agreement with the previous measurements~\cite{Bonvicini:2000qm,Dash:2017heu,Aaij:2015fua} and it is compatible with no $C\!P$ violation at the level of 2.4 standard deviations.

The second result concerns the search for $C\!P$ violation in $D_{(s)}^+ \to h^+ \pi^0$ and $D_{(s)}^+ \to h^+ \eta$ decays, where $h^+$ is $\pi^+$ or $K^+$,that provides interesting tests of the SM.
In particular, the two different weak phases contributing to the $D_s^+ \to K^+ \pi^0$, $D^+ \to \pi^+ \eta$ and $D_s^+ \to K^+ \eta$ decays allow $C\!P$ violation of the order of $10^{-3}$--$10^{-4}$ at tree-level, according to the SM~\cite{Cheng:2012wr}.
The result of the measurement of the $C\!P$ asymmetries in the $D_{(s)}^+ \to h^+ \pi^0$ and $D_{(s)}^+ \to h^+ \eta$ decays (excluding $D_s^+ \to \pi^+ \pi^0$, that is highly suppressed) performed by LHCb with an integrated luminosity of $9~\mathrm{fb}^{-1}$ and $6~\mathrm{fb}^{-1}$, respectively, is
\begin{alignat*}{7}
       \mathcal{A}_{C\!P}(D^+ \to \pi^+\pi^0) 	&= (-&&1.3 &&\pm 0.9 &&\pm 0.6 &)\%, \\
       \mathcal{A}_{C\!P}(D^+ \to K^+\pi^0) 	&= (-&&3.2 &&\pm 4.7 &&\pm 2.1 &)\%, \\
       \mathcal{A}_{C\!P}(D^+ \to \pi^+\eta)   &= (-&&0.2 &&\pm 0.8 &&\pm 0.4 &)\%, \\
       \mathcal{A}_{C\!P}(D^+ \to K^+\eta) 	&= (-&&6 &&\pm 10 &&\pm 4 &)\%, \\
       \mathcal{A}_{C\!P}(D^+_s \to K^+\pi^0) 	&= (-&&0.8 &&\pm 3.9 &&\pm 1.2 &)\%, \\
       \mathcal{A}_{C\!P}(D^+_s \to \pi^+\eta)  &= (&&0.8 &&\pm 0.7 &&\pm 0.5 &)\%, \\
       \mathcal{A}_{C\!P}(D^+_s \to K^+\eta)   &= (&&0.9 &&\pm 3.7 &&\pm 1.1 &)\%,
\end{alignat*}
where the first uncertainty is statistical and the second systematic~\cite{Aaij:2021wfw}.
All of the results are consistent with previous determinations~\cite{Babu:2017bjn,Guan:2021xer} and with no $C\!P$ violation, and the first five constitute the most precise measurements to date of the corresponding observables.

\subsection{Search for Time-Dependent $C\!P$ Violation in $D^0 \to K^+ K^-$ and $D^0 \to \pi^+ \pi^-$ Decays}

\begin{figure}[h]
\centering
\includegraphics[width=0.3\textwidth]{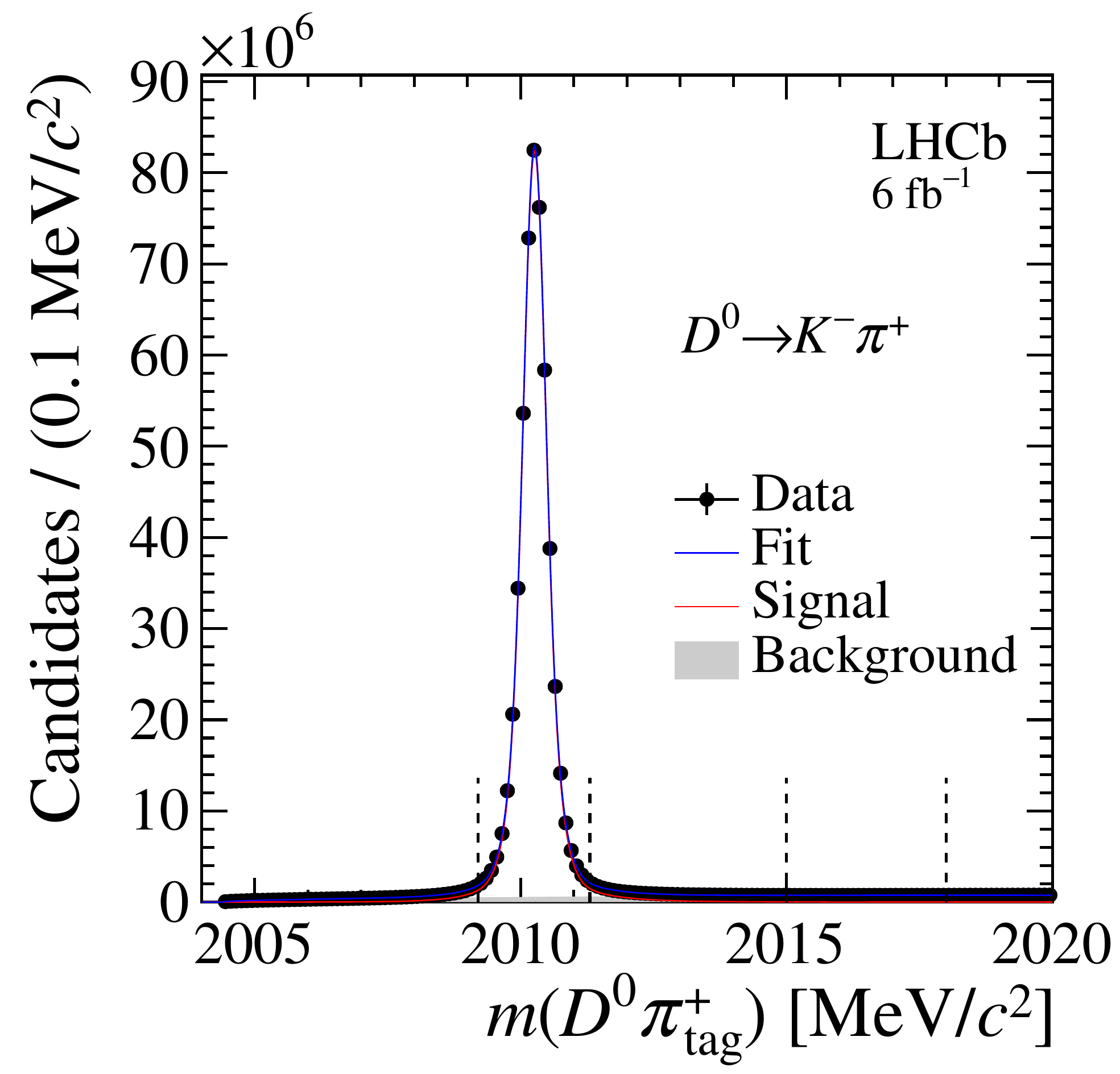}
\includegraphics[width=0.3\textwidth]{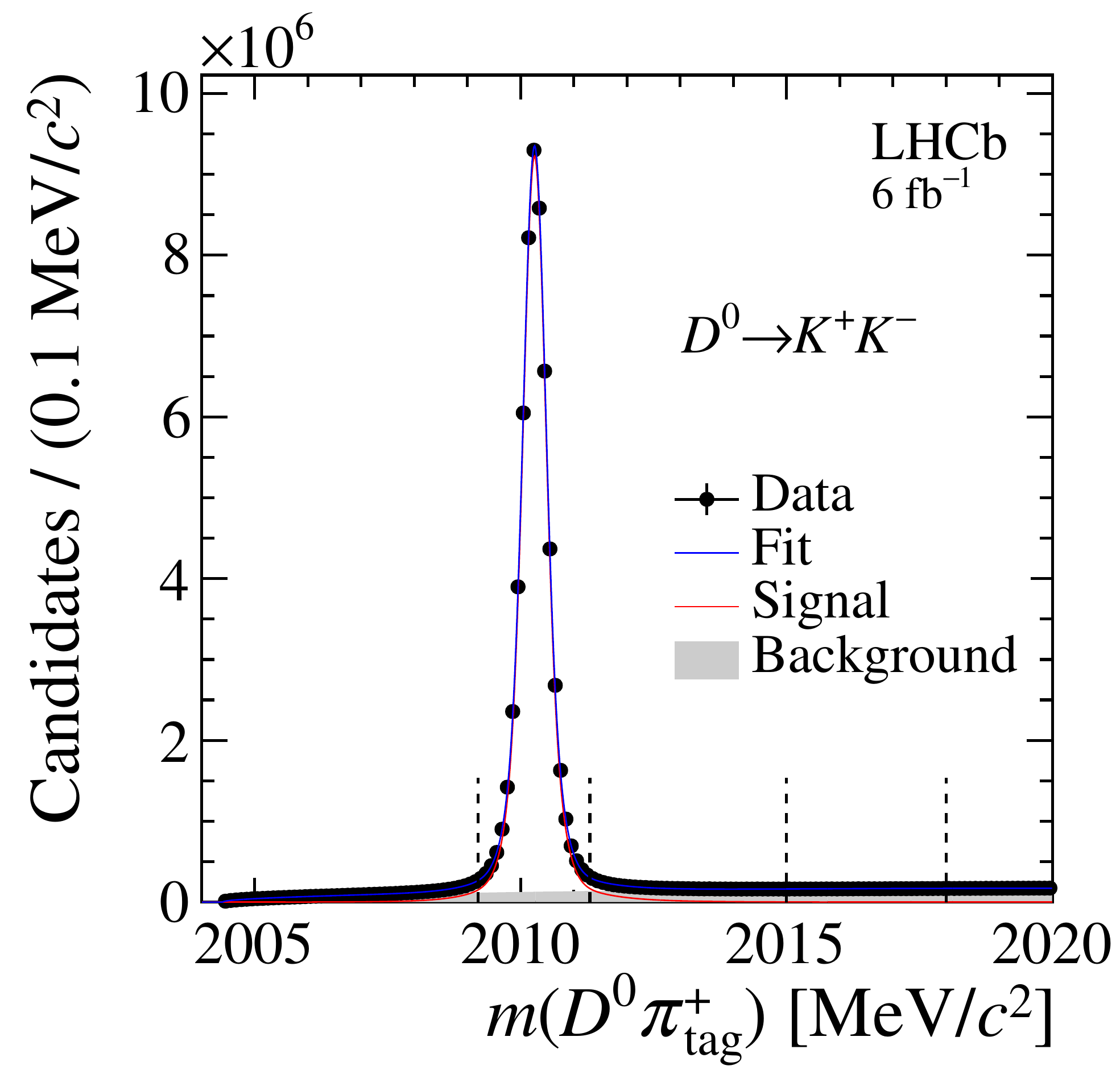}
\includegraphics[width=0.3\textwidth]{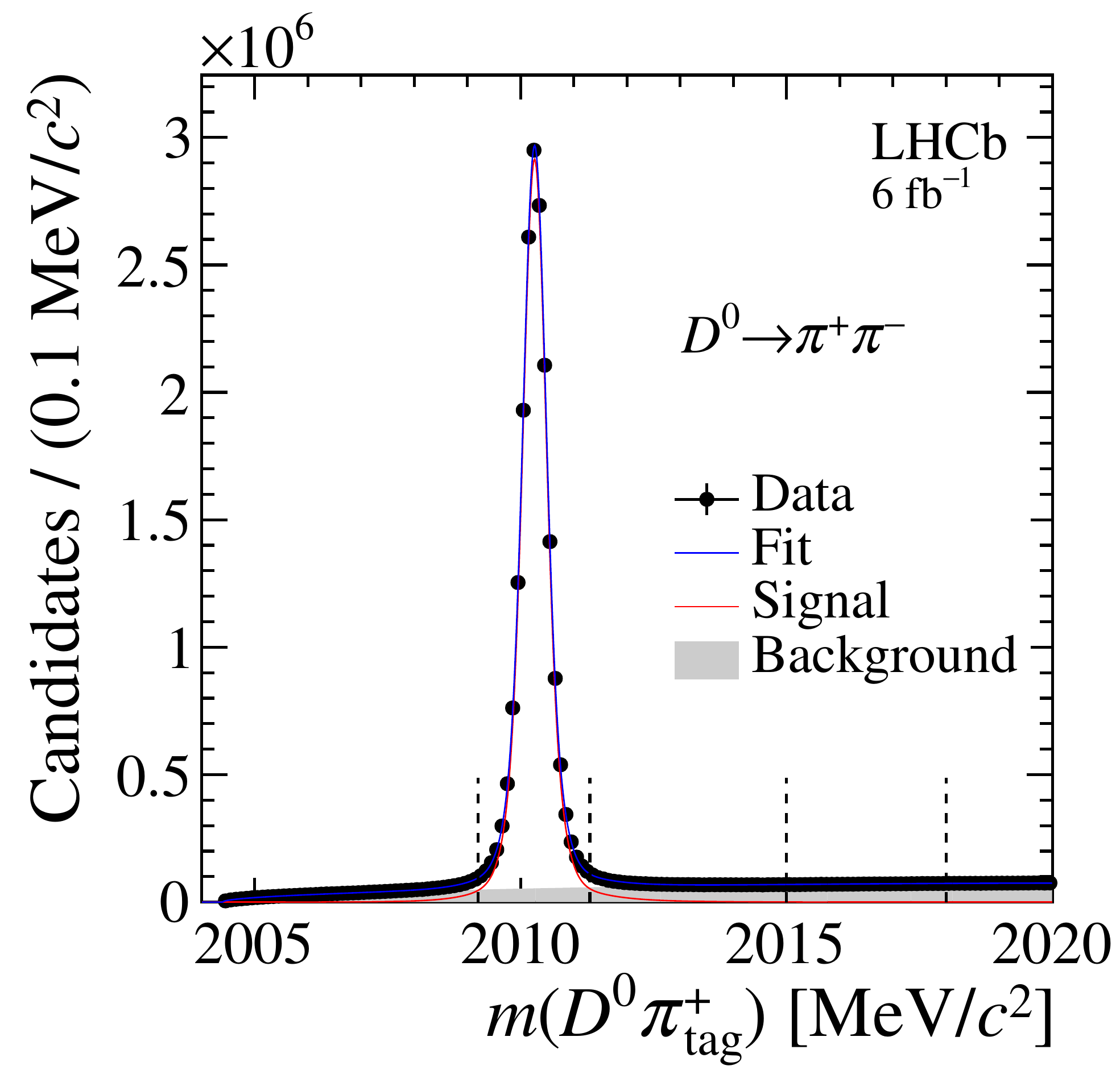}
\caption{Distribution of $m(D^0 \pi^+)$ for (left) $D^0 \to K^- \pi^+$, (center) $D^0 \to K^+ K^-$ and (right) $D^0 \to \pi^+ \pi^-$ candidates. The vertical dashed lines delimit the signal window and the sideband window used to remove the combinatorial background. Fit projections are overlaid.}
\label{fit_mass_DY}
\end{figure}

The magnitude of the $\Delta Y_f$ parameter is expected to be about $2 \times 10^{-5}$ for the final states $f=K^+K^-,\pi^+\pi^-$~\cite{Kagan:2020vri,Li:2020xrz}.
LHCb has performed a measurement of $\Delta Y_f$ with data collected during the LHC Run~2, corresponding to an integrated luminosity of $6~\mathrm{fb}^{-1}$, using $D^{*+} \to D^0 \pi^+$ decays originated from primary proton-proton interactions, where the $D^0 \to K^+ K^-$, $D^0 \to \pi^+ \pi^-$ and $D^0 \to K^- \pi^+$ decays are reconstructed.
The $C\!P$ asymmetry of  the $D^0 \to K^- \pi^+$ decay is known to be smaller than the current experimental uncertainty, so this mode is used as a control sample to validate the analysis.
The data sample is divided in 21 intervals of $D^0$ decay times in the range 0.45--8~$\tau_{D^0}$, where $\tau_{D^0}$ is the $D^0$ lifetime~\cite{Zyla:2020zbs}.
The signal yield in each decay-time interval is obtained by means of background subtraction in the $m(D^0 \pi^+)$ distribution. 
The flavour of the $D^0$ candidate at production is determined by the charge of the accompanying pion in the $D^{*+} \to D^0 \pi^+$ decay.
The weight assigned to the background candidates is determined with a binned maximum-likelihood fit to the $m(D^0 \pi^+)$ distribution, whose results are shown in Figure~\ref{fit_mass_DY}.

The selection requirements introduce a correlation between the kinematic variables and the $D^0$ decay time, resulting in an indirect time dependence of the production and detection asymmetries.
The nuisance asymmetries are therefore removed by equalising the kinematics of tagging $\pi^+$ and $\pi^-$ and of $D^0$ and $\overline{D}^0$ candidates, by weighting their kinematic distributions to their average.
Although the decays of $D^{*+}$ mesons originating from $B$ mesons are suppressed by a requirement on the impact parameter (IP) of the $D^0$ meson, some contamination is still present in the sample, resulting in a possible bias in the measured time-dependent asymmetry.
The effect of these secondary decays on the final measurement is corrected for by determining the size and asymmetry of this background by means of a binned maximum-likelihood fit to the bidimensional distribution of IP and decay time of the $D^0 \to K^- \pi^+$ candidates.
Systematic uncertainties are assessed related to charge-dependent biases on the reconstruction of the $D^{*+}$ mass, background removal, correction due to secondary decays, kinematic weighting and presence of misidentified background peaking in $m(D^0 \pi^+)$ distribution.

\begin{figure}
\centering
\includegraphics[width=0.45\textwidth]{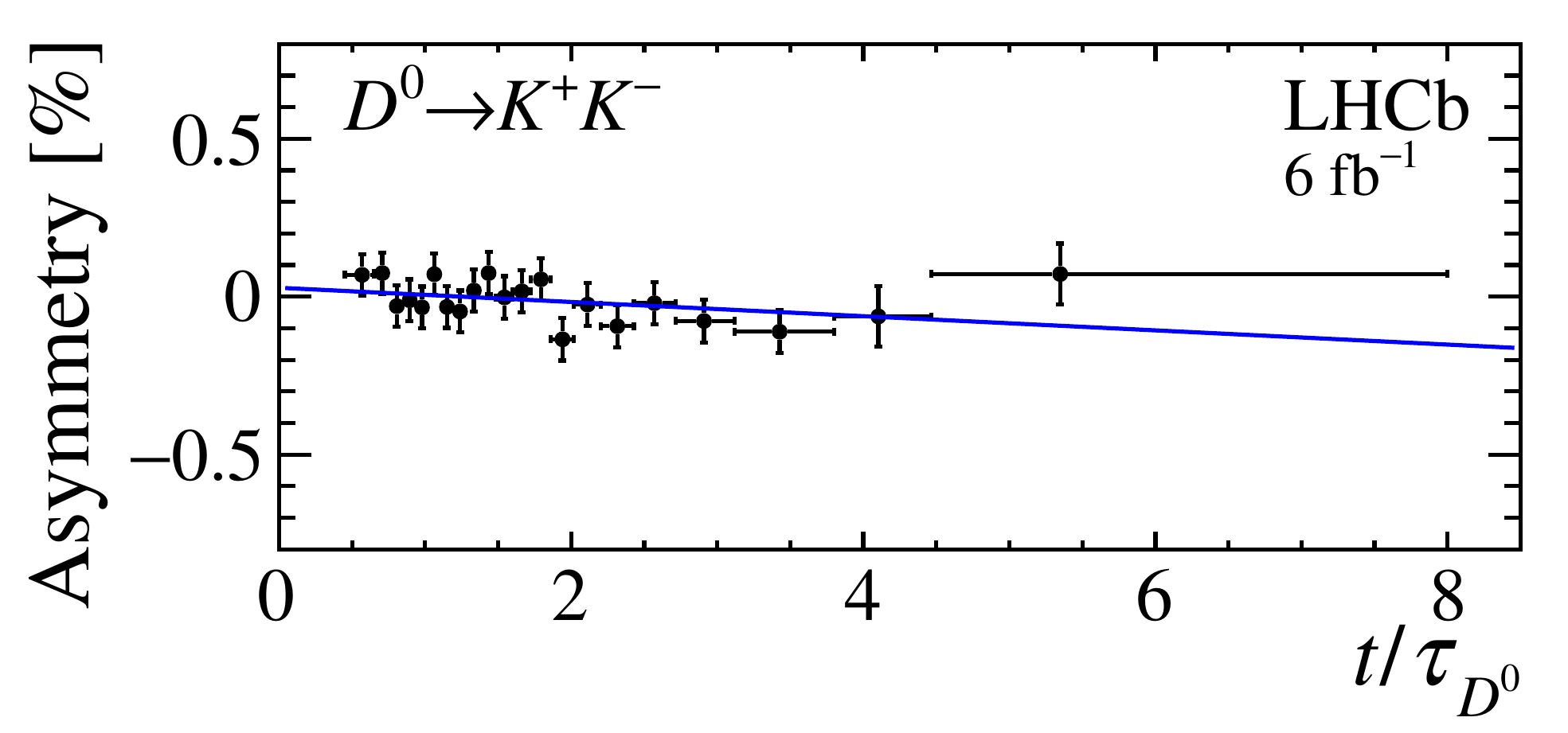}
\includegraphics[width=0.45\textwidth]{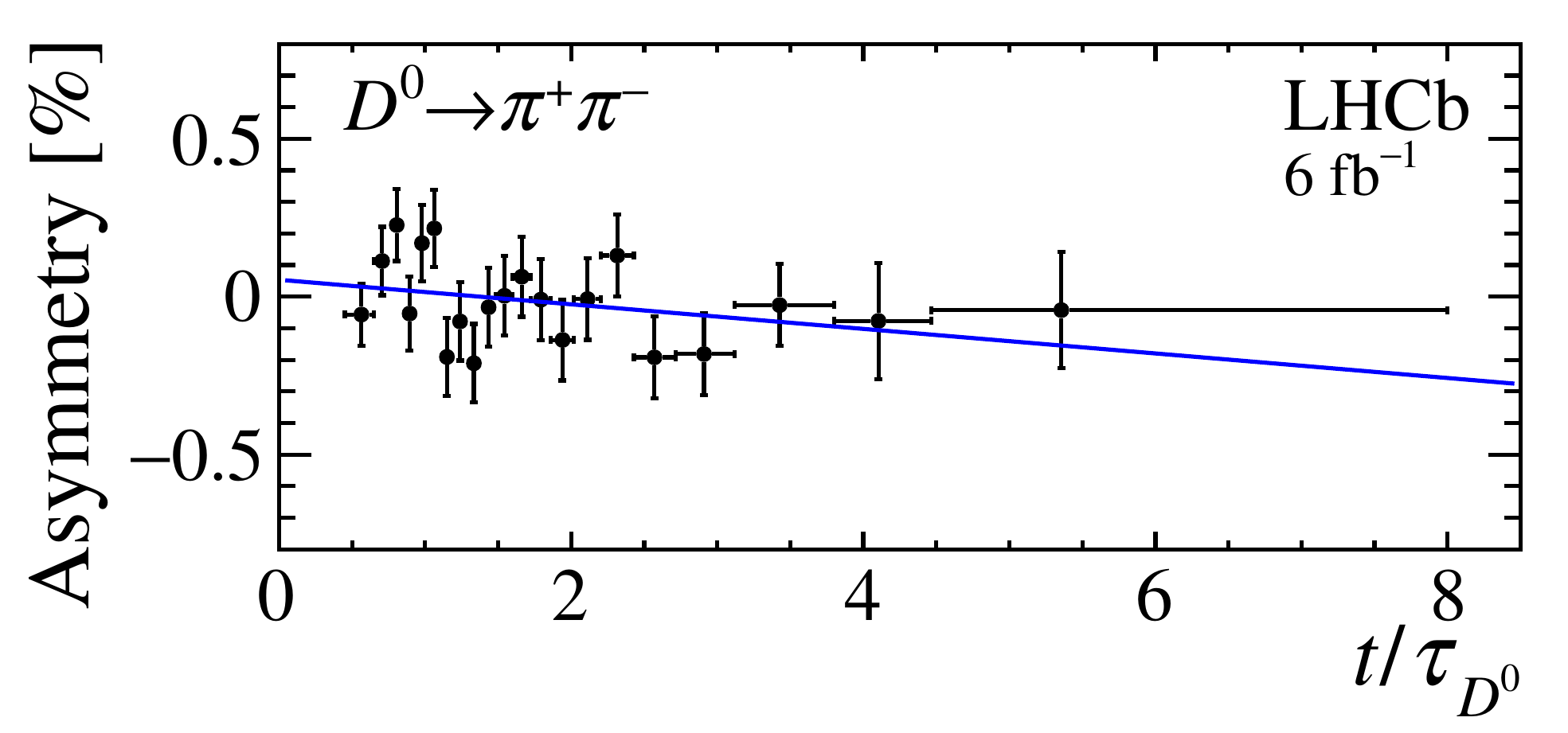}
\caption{Asymmetry as a function of $D^0$ decay time, for the (left) $D^0 \to K^+ K^-$ and (right) ${D^0 \to \pi^+ \pi^-}$ samples, with linear fit superimposed.}
\label{fit_KK_pipi_DY}
\end{figure}

{\sloppy The slope of the time-dependent asymmetry of the control sample is measured to be \mbox{$(-0.4 \pm 0.5 \pm 0.2) \times 10^{-4}$}, compatible with 0 as expected.
The time-dependent asymmetries of the $D^0 \to K^+ K^-$ and $D^0 \to \pi^+ \pi^-$ channels are shown in Figure~\ref{fit_KK_pipi_DY}, and the resulting slopes are
\begin{align*}
    \Delta Y_{K^+ K^-} &= (-2.3 \pm 1.5 \pm 0.3) \times 10^{-4}, \\
    \Delta Y_{\pi^+ \pi^-} &= (-4.0 \pm 2.8 \pm 0.4) \times 10^{-4},
\end{align*}
where the first uncertainties are statistical and the second are systematic~\cite{Lees:2012qh,Aaltonen:2014efa,Staric:2015sta,Aaij:2015yda,Aaij:2017idz,Aaij:2019yas}.
The values are compatible with each other and with previous determinations.
Assuming no final-state dependency, and taking into account the correlation between the systematic uncertainties, the combination of the two values is
\begin{equation*}
    \Delta Y = (-2.7 \pm 1.3 \pm 0.3) \times 10^{-4},
\end{equation*}
and is consistent with zero within two standard deviations~\cite{Aaij:2021pyl}.
The combination with previous LHCb measurements~\cite{Aaij:2015yda,Aaij:2017idz,Aaij:2019yas} leads to
\begin{align*}
    \Delta Y &= (-1.0 \pm 1.1 \pm 0.3) \times 10^{-4}.
\end{align*}
} 
This result, which is consistent with no time-dependent $C\!P$ violation, improves by nearly a factor of two the precision of the previous world average of $\Delta Y$~\cite{Amhis:2019ckw}.

\subsection{Observation of the Mass Difference between Neutral Charm-Meson~Eigenstates}

\begin{figure}
\centering
\includegraphics[width=0.45\textwidth]{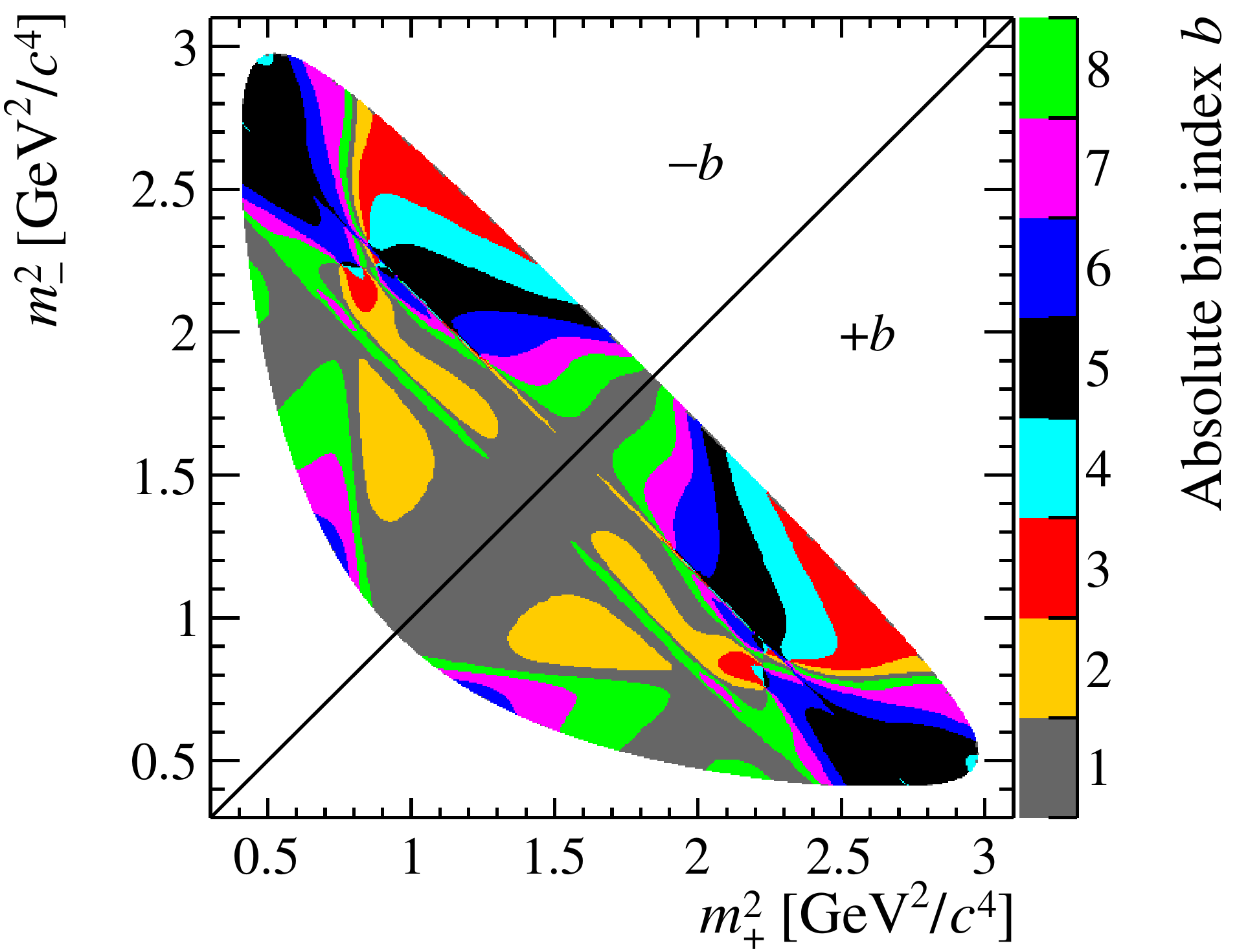}
\caption{Binning scheme of the $D^0 \to K_s^0 \pi^+ \pi^-$ Dalitz~plot~\cite{Libby:2010nu}.}
\label{KSpipi_dalitz_bins}
\end{figure}

The LHCb collaboration has recently measured the mixing and $C\!P$ violation parameters in $D^0 \to K_s^0 \pi^+ \pi^-$ decays with the data collected between 2016 and 2018, corresponding to $5.4~\mathrm{fb}^{-1}$ of integrated luminosity.
The analysis uses the ``bin-flip'' method~\cite{DiCanto:2018tsd}, which is a model-independent approach and does not require an accurate modelling of the efficiency.
The method consists in measuring, as a function of the $D^0$ decay time, the ratio of the number of decays between symmetric bins with respect to the bisector of the Dalitz plot defined by the two squared invariant masses $m^2(K_s^0 \pi^\pm)$.
The Dalitz bins, illustrated in Figure~\ref{KSpipi_dalitz_bins}, are defined in such a way that the strong-phase difference between the $D^0$ and $\overline{D}^0$ amplitudes within each bin is nearly constant~\cite{Libby:2010nu}.
The ratio of the number of decays between symmetric bins, determined for each decay-time interval and each $D^0$ flavour, is a function of the parameters $x$, $y$, $\phi$ and $|p/q|$.

The flavour of the $D^0$ candidate is determined by the charge of the accompanying pion in the reconstructed $D^{*+} \to D^0 \pi^+$ decay.
The signal yields are determined by means of a fit to the invariant mass difference $\Delta m \equiv m(D^{*+}) - m(D^0)$ in each Dalitz-plot bin and decay-time interval, separately for each $D^0$ flavour.
The signal selection includes requirements that introduce efficiency variations correlated between the phase-space coordinates and the $D^0$ decay time, resulting in a possible bias on the measurement.
The correlation is removed by applying a data-driven correction that makes the decay-time acceptance uniform in the phase space.
A further correction is performed to cancel detection asymmetries of the pions produced in the $D^0$ decays, since they depend on their kinematics and therefore on the Dalitz-plot coordinate: the two-track $\pi^+ \pi^-$ detection asymmetry is evaluated on the control samples $D^+_s \to \pi^+ \pi^+ \pi^-$ and $D^+_s \to \phi \pi^+$ .

\begin{figure}
\centering
\includegraphics[width=0.45\textwidth]{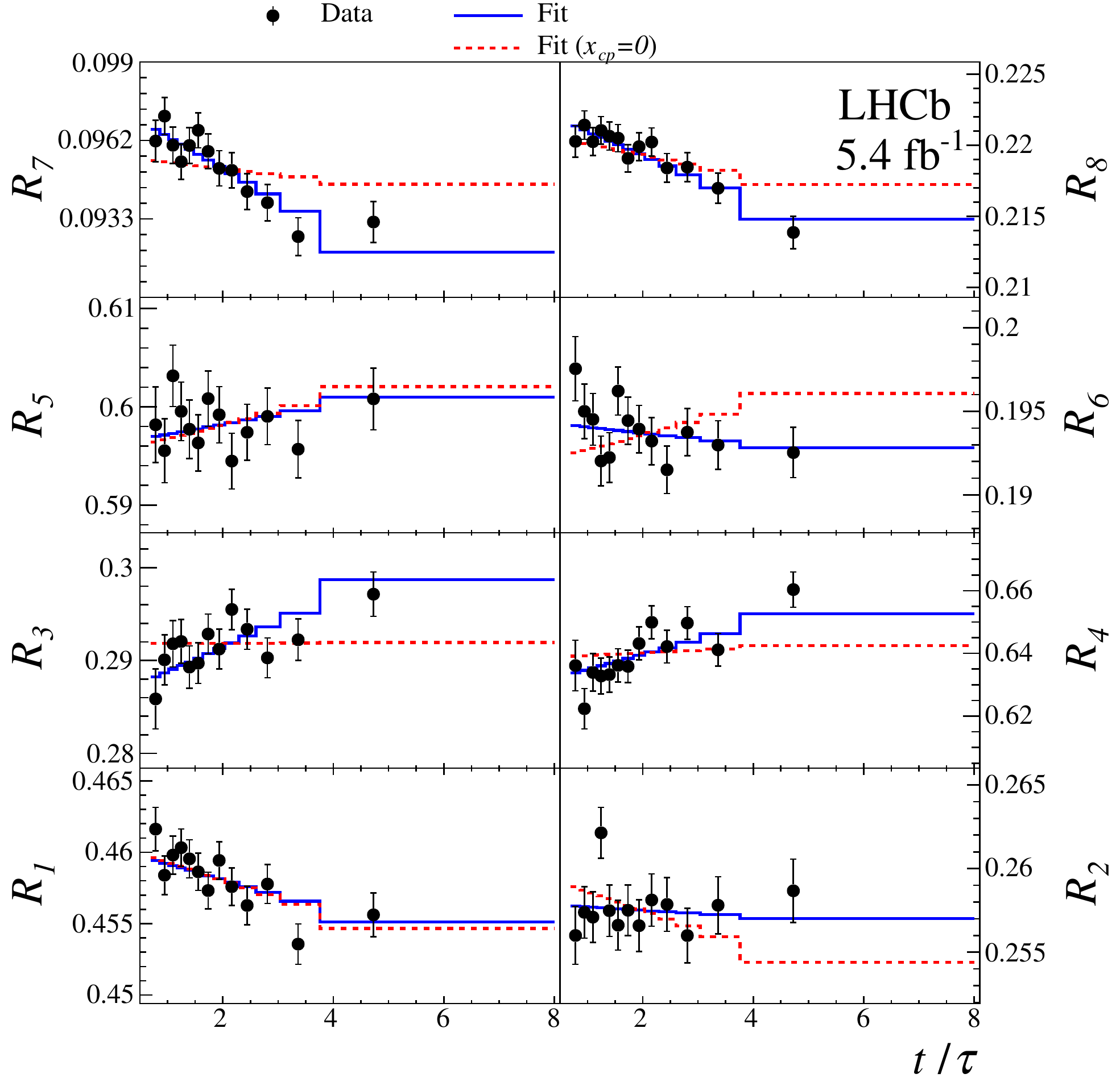}
\includegraphics[width=0.45\textwidth]{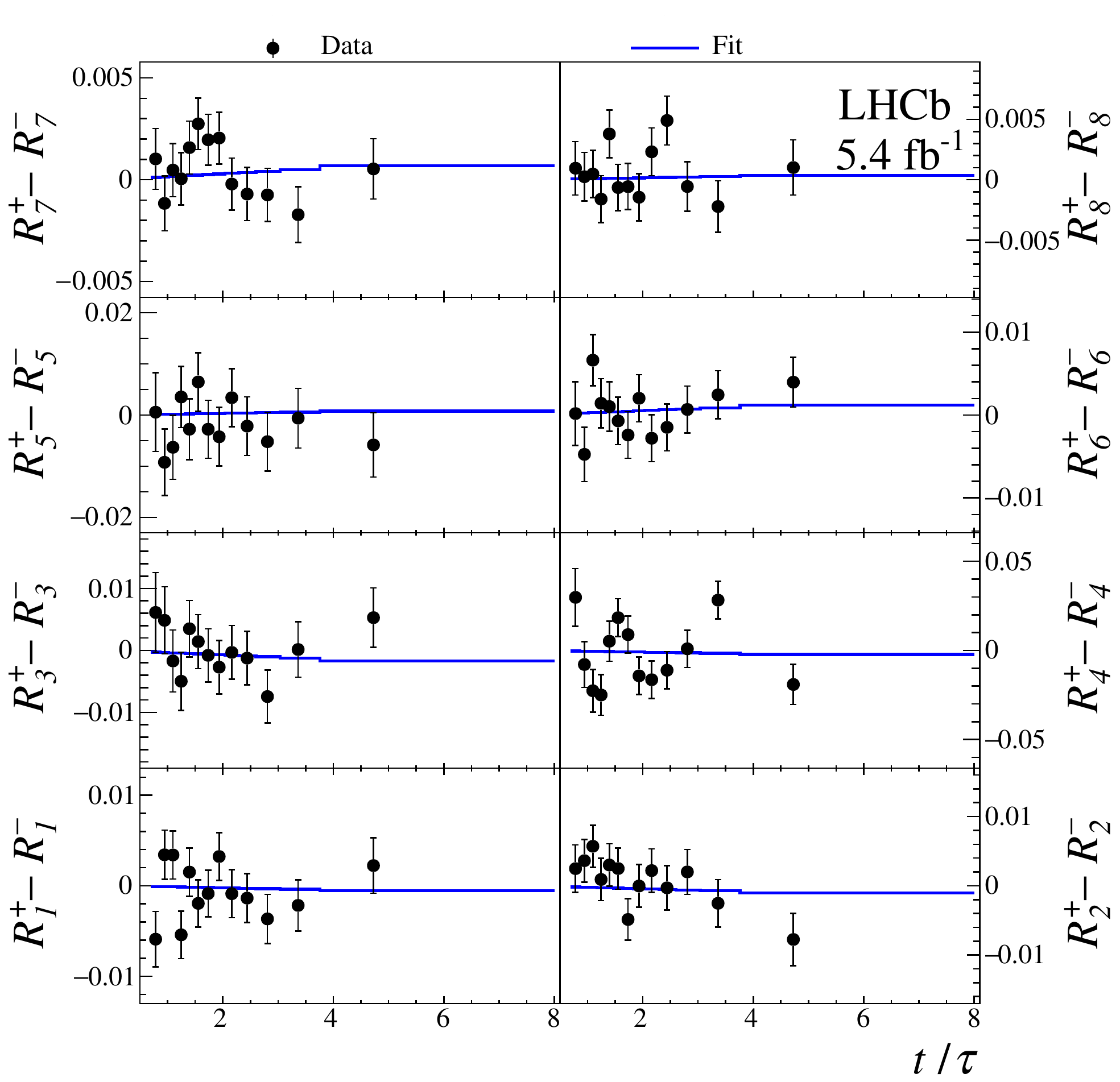}
\caption{(Left) $C\!P$-averaged yield ratios and (right) differences of $D^0$ and $\overline{D}^0$ yield ratios as a function of $D^0$ decay time, shown for each Dalitz-plot bin. Fit projections are overlaid.}
\label{KSpipi_fits}
\end{figure}

A fit is performed on all the corrected signal yield ratios to determine the mixing and $C\!P$ violation parameters.
The result is illustrated in Figure~\ref{KSpipi_fits}.
Various sources of systematic uncertainties are considered and assessed from ensembles of pseudoexperiments, to take into account contributions due to reconstruction and selection effects, presence of secondary $D^{*+}$ decays, decay-time and $m_\pm$ resolution, $\pi^+\pi^-$ detection asymmetry, fit model and to the approximation that strong-phase differences are constant within each bin.

A likelihood function of $x$, $y$, $|q/p|$ and $\phi$ is built from the results using a likelihood-ratio ordering that assumes the observed correlations to be independent of the true parameter values~\cite{Aaij:2013zfa}, whose best fit point is
\begin{alignat*}{2}
    x &= (3.98^{+0.56}_{-0.54}) \times 10^{-3}, &\quad\quad |q/p| &= 0.996 \pm 0.052, \\
    y &= (4.6^{+1.5}_{-1.4}) \times 10^{-3}, &\quad\quad \phi &= 0.056^{+0.047}_{-0.051}~\mathrm{rad}.\\
\end{alignat*}
This is the first observation of a non-zero value of the mass difference $x$ of neutral charm meson mass eigenstates, with a significance of more than seven standard deviations.
This result significantly improves limits on $C\!P$ violation in mixing in the charm sector.

\bibliographystyle{JHEP}
\bibliography{my_bib}

\end{document}